\documentclass[aps,prl,reprint,superscriptaddress]{revtex4-1}

\usepackage{color}
\usepackage{graphicx}
\usepackage{dcolumn}
\usepackage{bm}
\usepackage[dvipdfmx, colorlinks=true,citecolor=blue,linkcolor=blue]{hyperref}
\usepackage[mathlines]{lineno}

\begin{document}


\title{Effect of magnetism on lattice dynamics of {SrFe$_2$As$_2$}
  using high-resolution inelastic x-ray scattering}


\author{N. Murai}
\affiliation{Materials Dynamics Laboratory, RIKEN SPring-8 Center, 1-1-1 Kouto, Sayo, Hyogo 679-5148, Japan}
\affiliation{Department of Physics, Osaka University, 1-1 Machikaneyama-cho, Toyonaka,Osaka 560-0043, Japan}
\author{T. Fukuda}
\affiliation{Quantum Beam Science Center, Japan Atomic Energy Agency (SPring-8/JAEA), 1-1-1 Kouto, Sayo, Hyogo 679-5148, Japan}
\affiliation{Materials Dynamics Laboratory, RIKEN SPring-8 Center, 1-1-1 Kouto, Sayo, Hyogo 679-5148, Japan}
\author{T. Kobayashi}
\affiliation{Department of Physics, Osaka University, 1-1 Machikaneyama-cho, Toyonaka,Osaka 560-0043, Japan}
\author{M. Nakajima}
\affiliation{Department of Physics, Osaka University, 1-1 Machikaneyama-cho, Toyonaka,Osaka 560-0043, Japan}
\author{H. Uchiyama}
\affiliation{Research and Utilization Division, Japan Synchrotron Radiation Research Institute (SPring-8/JASRI),
1-1-1 Kouto, Sayo, Hyogo 679-5198, Japan}
\affiliation{Materials Dynamics Laboratory, RIKEN SPring-8 Center, 1-1-1 Kouto, Sayo, Hyogo 679-5148, Japan}
\author{D. Ishikawa}
\affiliation{Research and Utilization Division, Japan Synchrotron Radiation Research Institute (SPring-8/JASRI),
1-1-1 Kouto, Sayo, Hyogo 679-5198, Japan}
\affiliation{Materials Dynamics Laboratory, RIKEN SPring-8 Center, 1-1-1 Kouto, Sayo, Hyogo 679-5148, Japan}
\author{S. Tsutsui}
\affiliation{Research and Utilization Division, Japan Synchrotron Radiation Research Institute (SPring-8/JASRI),
1-1-1 Kouto, Sayo, Hyogo 679-5198, Japan}
\author{H. Nakamura}
\affiliation{Center for Computational Science and e-Systems, Japan Atomic Energy Agency, 178-4-4 Wakashiba, Kashiwa 277-0871, Japan}
\author{M. Machida}
\affiliation{Center for Computational Science and e-Systems, Japan Atomic Energy Agency, 178-4-4 Wakashiba, Kashiwa 277-0871, Japan} 
\author{S. Miyasaka}
\affiliation{Department of Physics, Osaka University, 1-1 Machikaneyama-cho, Toyonaka,Osaka 560-0043, Japan}
\author{S. Tajima}
\affiliation{Department of Physics, Osaka University, 1-1 Machikaneyama-cho, Toyonaka,Osaka 560-0043, Japan}
\author{A. Q. R. Baron}
\affiliation{Materials Dynamics Laboratory, RIKEN SPring-8 Center, 1-1-1 Kouto, Sayo, Hyogo 679-5148, Japan}
\affiliation{Department of Physics, Osaka University, 1-1 Machikaneyama-cho, Toyonaka,Osaka 560-0043, Japan}


\date{\today}

\begin{abstract}
  Phonon spectra of detwinned {SrFe$_2$As$_2$} crystals, as measured by
  inelastic x-ray scattering, show 
  clear anisotropy accompanying the magneto-structural transition at 200 K.
  We model the mode splitting using 
  magnetic DFT calculations, including a phenomenological reduction in force-constant
  anisotropy that can be 
  attributed to magnetic fluctuations. This serves as a starting point for
  a general model of phonons in 
  this material applicable to both
  the antiferromagnetically ordered phase and the paramagnetic phase. 
Using this model, the measured splitting in the magnetic phase below $\it T_{N}$,
and the measured phonon linewidth, 
we set a lower bound on the mean magnetic fluctuation frequency above $\it T_{N}$ at 210 K. 
\end{abstract}

\pacs{}

\maketitle


\begin{figure*}[bht]
\begin{center}
\includegraphics[width=17.50cm]{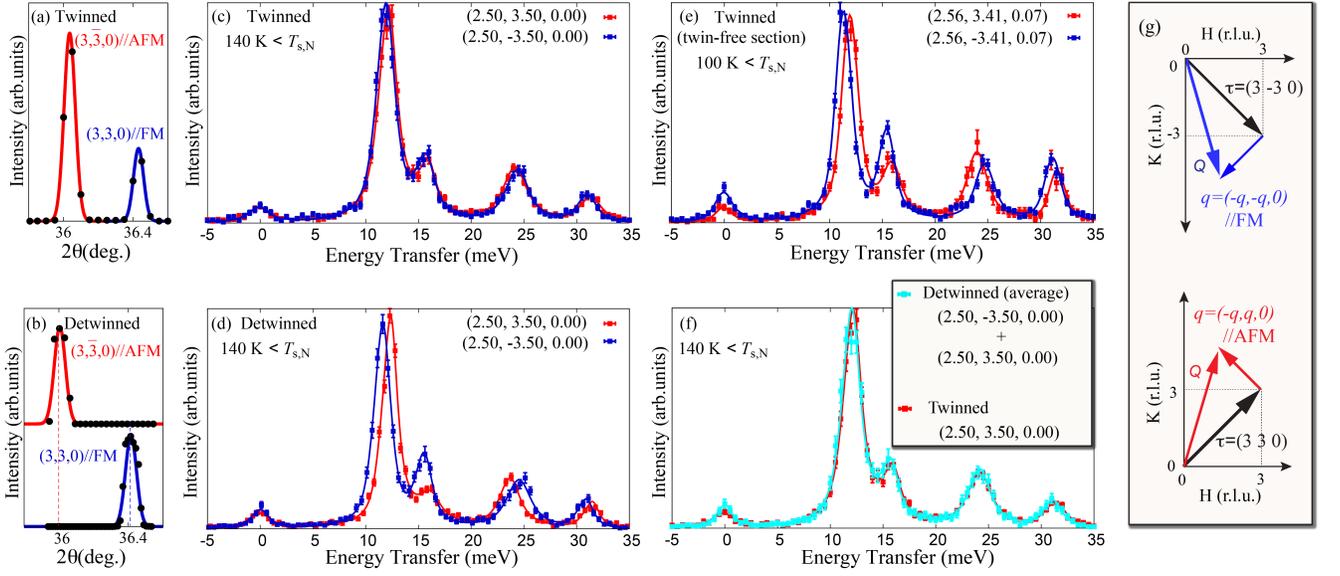}
\caption{(Color online) Anisotropy of phonon structure of detwinned SrFe$_{2}$As$_{2}$. 
(a) and (b), Typical 2$\theta$ scan of the tetragonal (3 3 0) and (3 $\overline{3}$ 0) 
reflections for twinned and detwinned crystals below $\it T_{s,N}$. 
(c) and (d), IXS spectra for twinned and detwinned crystals at \( { \bm q} = (-0.5, \pm 0.5, 0)  \) 
measured below $\it T_{s,N}$. Solid lines are fits to the data.
(e) IXS spectra measured at relatively twin-free section of twinned crystal. 
(f) Comparison of IXS spectra measured on the twinned crystal 
and a superposition of two IXS spectra measured on the detwinned crystal. 
(g) Schematic showing the directions in reciprocal space
where the IXS scans were measured.}
\label{Fig1}
\end{center}
\end{figure*}

\begin{figure}[thb]
\begin{center}
\includegraphics[width= 8.200cm]{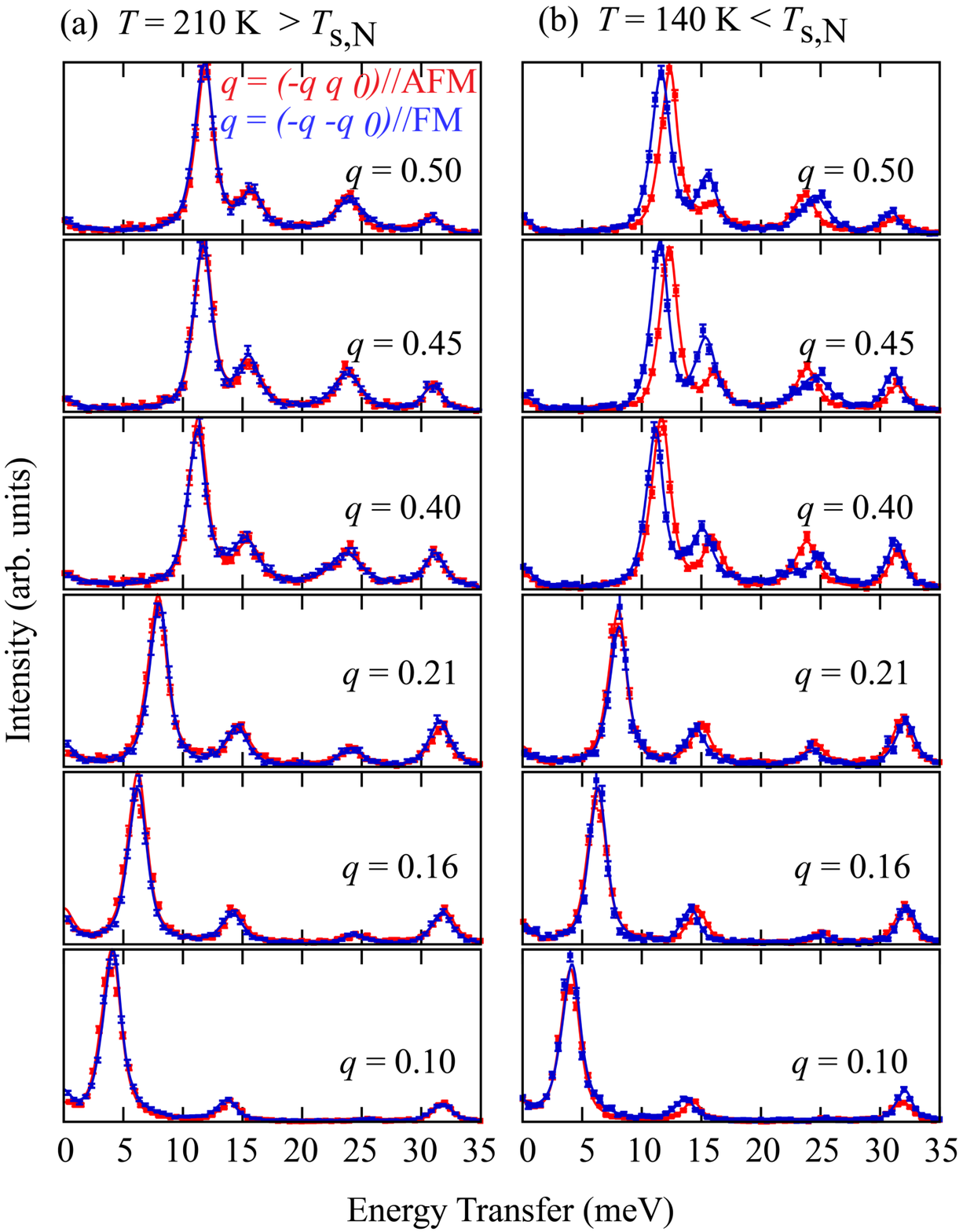}
\caption{(Color online) (a),(b) Temperature dependence of IXS spectra of detwinned {SrFe$_2$As$_2$} 
at \( {\bm Q} = (3 - q, 3 + q, 0) \) (red) and \( {\bm Q} = (3 - q, -3 - q, 0) \) (blue) 
corresponding to the two $\Gamma$-M directions. }
\label{Fig2}
\end{center}
\end{figure}

The close proximity of superconductivity to an antiferromagnetic (AFM) phase in the iron-pnictides 
suggests that magnetic fluctuations are involved in the pairing mechanism that leads to the high 
superconducting transition temperature ($\it T_{c}$)\cite{Y.Kamihara}. In fact, early density
functional theory (DFT) calculations suggested that the electron-phonon coupling is too weak 
to account for the observed high-$\it T_{c}$\cite{L.Boeri}, implying that the superconductivity is not 
phonon mediated. On the other hand, the physical properties of iron-pnictides do exhibit a strong sensitivity 
to the lattice\cite{A.Kreyssig,C.-H.Lee,Y.Mizuguchi,T.Yildirim,K.Kuroki}. 
This makes it interesting to study the relation between spin and lattice degrees of freedom in iron-pnictides.

Members of the {AFe$_2$As$_2$} (A = Ba, Sr or Ca) iron-pnictide family typically exhibit, on cooling, 
a tetragonal ($\it I{\rm 4}/mmm$) to orthorhombic ($\it Fmmm$) structural phase transition below $\it T_s$ 
followed by a magnetic phase transition into a collinear AFM ordered phase below 
$\it T_N$ (\(\leq\) $\it T_{s}$)\cite{J.Zhao_PRB2008, Q.Huang, M.Rotter}, both of which break the
90$^{\circ}$ rotational C$_{4}$ symmetry of the 
underlying tetragonal lattice. The emergence of the symmetry breaking also manifests in pronounced 
in-plane anisotropies as reported by transport\cite{J.-H.Chu}, angle-resolved photoemission 
spectroscopy (ARPES)\cite{M.Yi}, neutron scattering\cite{J.Zhao_nat.phys2009},
optical spectroscopy\cite{M.Nakajima} 
and torque magnetometry\cite{S.Kasahara}. This is often referred to as nematic order, and its origin 
has been one of the most intensively debated issues in iron-pnictide materials\cite{R.M.Fernandes}. 
Despite the evidence of anisotropic behaviour, the phonon response is surprisingly 
isotropic\cite{T.Fukuda,D.Reznik,D.Parshall1}. Phonon anisotropy should exist, in principle,
and mode splitting has been seen using Raman scattering\cite{L.Chauviere}, but anisotropy has
not been observed at non-zero momentum transfers. 

Here we report an inelastic x-ray scattering (IXS) study of detwinned single crystals of 
{SrFe$_2$As$_2$}. Our results clearly show anisotropy in phonon structure below $\it T_{s} = \it T_{N}$ 
\cite{J.Zhao_PRB2008} characterized by energy shifts and intensity changes of phonons
at tetragonally-equivalent momentum transfers. To the best of our knowledge,
this is the first observation of phonon anisotropy in 
iron-pnictides at finite momentum transfer. We compare our results to the DFT calculations 
and find that the best agreement is obtained by reducing the anisotropy of magnetic calculations 
by roughly a factor of 3. Based on this analysis, the underlying magnetic state of iron-pnictides
and its effect on phonon response are discussed.

Single crystals of {SrFe$_2$As$_2$} were grown by a self flux method described in Ref. \cite{T.Kobayashi}. 
The crystals undergo concomitant structural and magnetic phase transitions at 
$\it T_{s,N}$ = 200 K.
The tetragonal lattice parameters are \({\it a = b =} 3.924\) {\AA} and \({\it c } = 12.364\) {\AA} at room
temperature. 
Throughout this paper, we use tetragonal notation with axes along the 
next-nearest-neighbor iron atoms. 
The magnetic structure of {SrFe$_2$As$_2$} 
below $\it T_{N}$ is collinear with the ordered moment aligned antiferromagnetically (ferromagnetically) along 
the [1 $\overline{1}$ 0] ( [1 1 0] ) direction corresponding to the longer $\it a$-axis (shorter $\it b$-axis)
of the orthorhombic lattice.  

In the AFM phase, {SrFe$_2$As$_2$} generally forms small twin domains, which obscure its
intrinsic anisotropic properties. 
To avoid twinning, we applied uniaxial compressive pressure to a crystal before cooling below $\it T_{s,N}$.
The crystal was glued between two copper prongs (similar to a tuning fork)
with a screw used to carefully adjust the pressure parallel to one [1 1 0] direction. 
For a twinned crystal, Bragg reflections of the tetragonal \( {(h, h, 0)}\) type exhibit 
splitting in 2$\theta$ corresponding to two distinct $\it d$-spacings (Fig. \ref{Fig1} (a)). 
Application of uniaxial pressure favors formation of domains with the shorter lattice constant
parallel to the pressure axis [see Fig. \ref{Fig1}(b)].

Phonon measurements using IXS were performed at BL35XU\cite{A.Q.R.Baron1} and BL43LXU
\cite{A.Q.R.Baron2} of SPring-8 in Japan. 
The scattered radiation was collected using a two-dimensional (2-D) analyzer array on a 10 m horizontal 
2$\theta$ arm, which allows parallelization of data collection in a 2-D section of momentum space
\cite{A.Q.R.Baron3}. 
The energy resolution was determined from measurements of plexiglas to be 1.5 meV - 1.8 meV at 21.747 keV 
(Si (11 11 11) geometry) depending on the analyzer crystals.  
The data were collected in transverse geometries along two tetragonally-equivalent lines 
corresponding to the two $\Gamma$-M directions
\footnote{The literature is not always consistent as to the labelling of the \((0.5,0.5,0)\) Q point, often using M and occasionally X.
  We note this inconsistency here and use M in this paper.}: 
(1) \( {\bm Q} = (3 - q, 3 + q, 0) \) with \( {\bm q} = (-q, q, 0) \) parallel to the AFM ordered
direction; 
(2)  \( {\bm Q} = (3 - q, -3 - q, 0) \) with \( {\bm q} = (-q, -q, 0) \) 
parallel to the FM ordered direction [see Fig. \ref{Fig1} (g)]. 
These two $\Gamma$-M directions become inequivalent in the AFM phase. 
Quantitative results were obtained by fitting the IXS spectra with the sum of a 
resolution-limited elastic peak and several damped harmonic oscillators (DHOs) for the phonon modes 
convoluted with the experimentally determined resolution function.

In Figs. {\ref{Fig1} (c) and (d)}, we compare IXS spectra at \( {\bm q} = (-0.5, \pm0.5, 0)  \)
for twinned and detwinned {SrFe$_2$As$_2$} crystals in the AFM ordered state.
For the twinned crystal in Fig. \ref{Fig1} (c), there is no clear evidence for any change between two
tetragonally-equivalent momentum positions due to the twinning, except for a tiny ($\sim$ 0.1 meV)
energy shift of some of the modes. 
However, once detwinned, clear phonon anisotropy can be observed as easily seen in Fig. \ref{Fig1} (d) 
where both the frequencies and intensities of the modes change. 
In one case, we were able to observe the same effect even without the application of external pressure 
when we were fortunate enough to isolate large single domain section of a crystal. The resulting twin 
structure was not stable 
when the temperature was cycled, so most experiments were carried out under pressure.
However, the fact that the same effect was observed without pressure [see Fig. \ref{Fig1} (e)] serves to
confirm that the pressure does not significantly affect the response of these samples. 
Furthermore, IXS spectra measured on the stress-free twinned crystal can be reproduced by averaging 
those measured on detwinned crystal [see Fig.\ref{Fig1} (f)]. 
This further confirms the intrinsic anisotropy of phonon structure in the AFM phase.

The momentum dependence of the changes observed across $\it T_{s, N}$ are shown in Fig.\ref{Fig2}.  
Above $\it T_{s,N}$, IXS spectra are essentially identical in the two $\Gamma$-M directions 
[see Fig. \ref{Fig2} (a)], as the crystal has \(\it C_{\rm 4}\) rotational symmetry in 
tetragonal paramagnetic (PM) phase. In contrast, on lowering temperature below $\it T_{s,N}$,
which breaks the \(\it C_{\rm 4}\) rotational symmetry, anisotropic phonon shifts develop
between two $\Gamma$-M directions [see Fig.\ref{Fig2} (b)]. No significant change in
line-width was observed across $\it T_{s,N}$ ($\it e.g.,$ 
full-width at half-maximum (FWHM) of the 24 meV mode at $\it q$ = 0.50 is 0.70 $\pm$ 0.10 meV
for 140 K and 0.74 $\pm$ 0.09 meV for 210 K ). 
Note that the small orthorhombic structural distortion ((a - b)/(a + b) $\sim$ 0.5 $\%$) is 
expected to have only a very small direct effect on the phonons between the two $\Gamma$-M 
directions\cite{D. Reznik}. We therefore expect that the changes in the phonon energies are 
predominantly the result of the onset of the magnetic order, as opposed to the small orthorhombic 
structural distortion.

We compare the experimental results to the DFT calculations. All calculations were performed using 
the relaxed tetragonal structure ($\it I{\rm 4}/mmm$) with generalized gradient approximation (GGA) 
using projector-augmented wave (PAW) pseudopotentials, as implemented in the
Vienna Ab initio Simulation Package 
(VASP)\cite{G.Kresse1, G.Kresse2, G.Kresse3}. Phonons were calculated using a direct 
method\cite{K.Parlinski} for both nonmagnetic and magnetic ground states.

\begin{figure}[thb]
\begin{center}
\includegraphics[width= 8.80cm]{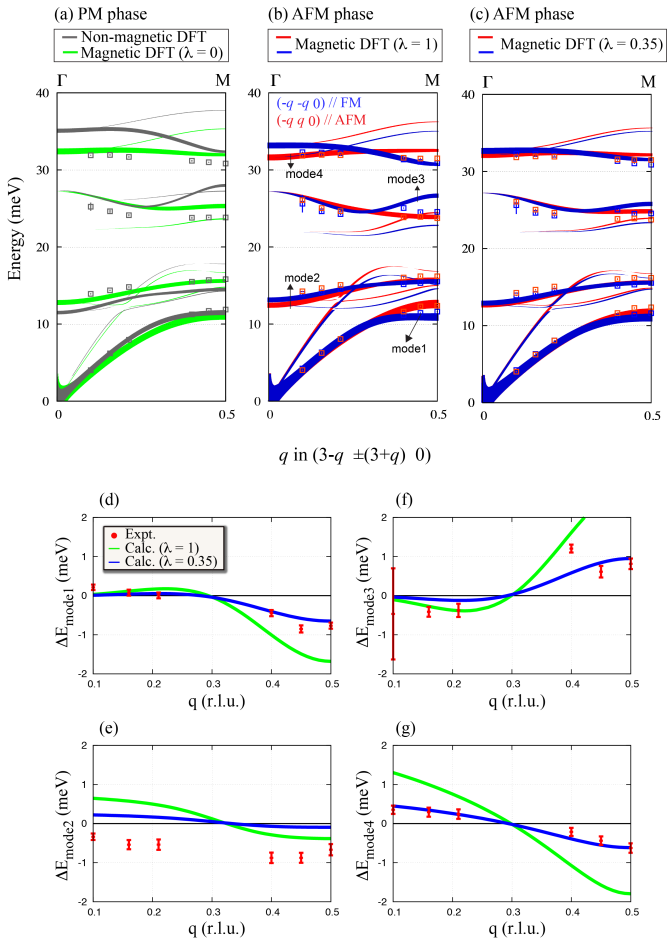}
\caption{(Color online) (a)-(c) Comparison of the measured dispersion for detwinned {SrFe$_2$As$_2$}  
and various DFT calculations at \( {\bm Q} = (3-q, \pm (3+q), 0) \). 
In (b) magnetic and (c) rescaled magnetic calculations, there are two inequivalent $\Gamma$-M directions
pointing along the AFM and the FM ordered directions, shown in red and blue, respectively.
Lines represent calculated phonon dispersions weighted by the structure factor, whereas the data points represent
experimental phonon energies.
(d)-(g) Comparison of the measured mode splitting for detwinned {SrFe$_2$As$_2$}  
and DFT calculations at \( {\bm Q} = (3-q, \pm (3+q), 0) \).
Mode 1 is the TA mode, modes 2, 3, and 4 have, respectively, \(E_{u}\), \(B_{1g}\), and \(E_{u}\) symmetry
at \(\Gamma\) in the tetragonal structure (\(I4/mmm\))} 
\label{Fig3}
\end{center}
\end{figure}

Fig. \ref{Fig3} shows the results of calculations and the data from the detwinned crystal. 
The non-magnetic calculation (grey curve in Fig. \ref{Fig3} (a)) fails to reproduce the experimental data, 
especially for the branch dispersing from $\sim$ 35 meV at $\Gamma$ point.  
The calculated energy of this branch is significantly higher than observed. 
The calculations can be brought into better agreement with the data if magnetism is included in 
the calculations [see the red and blue curves in Fig. \ref{Fig3} (b)], as suggested by the earlier work
\cite{T.Fukuda, D.Reznik, S.E.Hahn}. 
In general, magnetism has the biggest effect on some high energy branches, with lower energy branches
relatively unaffected. Note that the change in relative intensity for the modes at $\sim$ 15 meV in
Fig.\ref{Fig1} (d) 
is also reproduced by calculations. 
However, the magnetic calculations predict splitting of modes that is much larger than we observe.  
This is evident in Fig. \ref{Fig3} (b), where the magnetic calculations 
give mode splitting of several meV near zone boundary, while our data shows splitting of $\sim$ 1 meV at most.

To gain insight into our results, we consider a phenomenological 
modification to the real space force-constant (FC) matrices. 
We decompose the magnetic FC matrices into parts obeying $\it C_{\rm 4}$ (tetragonal) and $\it C_{\rm 2}$
(magnetic) 
rotational symmetry as 
$\phi_{d\alpha,d^{'}\beta} = \phi^{C_{4}}_{d\alpha,d^{'}\beta} + \phi^{C_{2}}_{d\alpha,d^{'}\beta}$ where 
$\alpha$ and $\beta$ are the cartesian directions, and $\it d$ and $d^{'}$ specify a pair of atoms. 
The symmetry-recovered $\it C_{\rm 4}$ matrices $\phi^{C_{4}}_{d\alpha,d^{'}\beta}$ are obtained by 
averaging tetragonally-equivelent matrices in the magnetic DFT. 
A similar symmetrization procedure has also been used to compute the FC matrices of iron in the 
high-temperature PM phase\cite{F.Kormann}.  We then reconstruct the effective FC matrices 
$\phi^{\rm eff}_{d\alpha,d^{'}\beta}$ by scaling $\it C_{\rm 2}$ term 
$\phi^{C_{2}}_{d\alpha,d^{'}\beta}$ linearly in $\lambda$, 
\begin{equation}
\label{eq:eq1}
\phi^{\rm eff}_{d\alpha,d^{'}\beta} = \phi^{C_{4}}_{d\alpha,d^{'}\beta} + \lambda\phi^{C_{2}}_{d\alpha,d^{'}\beta}
\end{equation}
where $\lambda$ is a scaling factor that accounts for renormalization of the FC anisotropy. 
The optimal value of $\lambda$ in the AFM phase is determined to be $0.35 \pm 0.05 $ by numerical 
optimization of the magnitude of mode splitting $\Delta$E to match the measured values at each 
momentum transfer. We also considered a model where $\phi^{C_{2}}_{d\alpha,d^{'}\beta}$ 
exponentially decays with length between a pair of atoms, but the best fit was obtained with the uniform 
linear scaling in Eq. \ref{eq:eq1}. This, as well as the fact that we observe similar mode splitting at both 
high and low $\bm q$ regions, indicates that there is not any characteristic length scale to the renormalization 
of FC anisotropy. 

In Fig. \ref{Fig3} (c), we compare the rescaled magnetic calculations with the experimental data. 
One can see that the rescaled calculations show better overall agreement. 
In Fig.\ref{Fig3} (d) - (g), the momentum dependence of the calculated mode splitting is shown in 
comparison with the experimental data. With a linear rescaling of FC anisotropy, the calculated mode splitting 
can be reduced to the level of the experimental data, except for mode 2 at $\sim$ 14 meV [see Fig.\ref{Fig3} (e)]. 
The discrepancy between the calculated dispersion for mode 2 and the experimental values might be an indication
of some missing ingredient that is not properly included in the calculations (e.g. orbital ordering).
Nonetheless, the degree of agreement with the experimental data suggest that linear rescaling
model in Eq. \ref{eq:eq1} is a good starting point to describe lattice dynamics of iron-pnictides. 

Having established the better overall agreement with the data, we now move on to the physical 
interpretation of our results. The overestimation of the phonon anisotropy by the DFT calculations is
reminiscent of the tendency of DFT calculations in iron-pnictides to give a significantly larger ordered moment
(\(\sim 2 {\mu}{\rm {_B}}\)/Fe)
\cite{I.I.Mazin_PRB2008, I.I.Mazin_Nat.Phys2009} than is observed in most
experiments (\(\sim 0.9 \mu_{\rm B}\))\cite{J.Zhao_PRB2008, Q.Huang}. 
It is interesting to note that a reduction factor of \({\lambda} = 0.35\) in Eq. (\ref{eq:eq1}) relative to the 
DFT is roughly comparable to that found for magnetic moment. This suggests that the magnitude of mode 
splitting is proportional to the size of the ordered moments.
On the other hand, recent Fe 3$\it s$ core level photoemission spectroscopy has 
revealed the presence of large local moment of \(\sim 2 {\mu_{\rm B}}\) fluctuating on a 
femtosecond time scale in the PM phase\cite{P.Vilmercati}. 
These fluctuating local moments are expected to be ordered below $\it T_{N}$,
but the size of the ordered moment is, as mentioned above, significantly smaller than that of the
local moments.
On the theoretical side, there have been several attempts to understand the origin of the reduced
ordered moment beyond DFT, using dynamical mean-field theory (DMFT)\cite{P.Hansmann, Z.P.Yin1, Z.P.Yin2}.
These can explain the presence of large local moments which only give rise to much smaller ordered
moment below $\it T_{N}$. For example, Z. P. Yin $\it et \ al.$ have suggested that there is the strong orbital
differentiation, with the $\it t_{2g}$ orbitals more correlated than the $\it e_{g}$ orbitals\cite{Z.P.Yin2}.
In this situation, the static ordered moment originates predominantly from more localized
$\it t_{2g}$ orbitals while fluctuating local moments in the $\it e_{g}$ orbitals
do not acquire a static component below $\it T_{N}$. 
Such orbital-selective correlations result in the reduced ordered moment in the AFM phase, and in analogy to
this, one can expect the reduced phonon anisotropy. 
In this context, the \(C_{\rm 2}\) term of Eq.(\ref{eq:eq1}) arises from the ordered moment while the \(C_{\rm 4}\)
term includes the contribution of the fluctuating moments.
Note that even in the PM phase, where long-range AFM order is destroyed,
the averaged magnetic FC matrices $\phi^{C_{4}}_{d\alpha,d^{'}\beta}$ (\(\lambda = 0\) in Eq.\ref{eq:eq1}) gives
better agreement with phonon dispersion than non-magnetic DFT [see grey and green curves in Fig.\ref{Fig3} (a)].  
We take this as an indication of the presence of fluctuating magnetism above $\it T_{N}$,
consistent with Ref. \cite{P.Vilmercati}.

To understand the effect of fluctuations on the phonon response, we consider a simple model of a mass,
$\it m$, on a spring, where the spring constant fluctuates between two values, \(k_{1}, k_{2}\) at random times
governed by a negative exponential distribution with mean dwell time $\tau$.  As shown in the supplemental
materials\footnote{See Supplemental Material for Phonon line-width and splitting in the
  presence of a fluctuating force constant.}, the shape of the power spectrum of the displacement is governed by $\tau$ and the
frequency difference \(s = (\sqrt{k_{2}/m} - \sqrt{k_{1}/m})/2\pi\)
: for slow fluctuations, \(\tau s > 1\), there are two well defined lines whose width
(FWHM) is given by \(\Gamma/h \sim (\pi \tau)^{-1} \), 
while for fast fluctuations, \(\tau s < 0.2\), the lines coalesce into a single line of width
\(\Gamma/h \sim \sqrt{8/\pi}\tau s^{2} \).
Taking, as an example, the phonon widths quoted earlier of \(\Gamma_{+} = 0.74 \pm 0.09\) meV
above $\it T_{s,N}$ and \(\Gamma_{-} = 0.70 \pm 0.10\) meV below $\it T_{s,N}$ with a splitting of
\(hs = 0.81\) meV, and assuming other contributions to the line width do not change through $\it T_{s,N}$,
suggests a mean magnetic fluctuation frequency \(1/\tau > 1.4\) THz.
This assumes the broadening above \(T_{s,N}\) is less than \( 0.04 + 0.14 =0.18\) meV,
where the 0.14 is the error on the difference. 
This frequency is lower than the limits suggested by other methods\cite{P.Vilmercati}, but still valuable. 
A measurement with higher resolution (e.g. $\sim$ 0.01 meV as has been demonstrated in Ref.\cite{P.Aynajian})
might determine the fluctuation frequency more exactly.

We note that recently it was found, for small $\bm q$, that TA modes polarized in the [1 0 0] direction
soften at \cite{J.L.Niedziela} and above \cite{D.Parshall2} $\it T_{s,N}$, and
this was suggested to be related to the size of the fluctuating magnetic domains\cite{D.Parshall2}.
While different than the present work, where we observe clear energy splitting over the full zone below
$\it T_{s,N}$, for differently polarized modes, that work also shows the sensitivity of the phonon measurements to
magnetic order, and, indeed serves to highlight the potential to use careful phonon measurements to
investigate both static and dynamical aspects of magneto-elastic coupling.

In summary, we reveal phonon anisotropy of SrFe$_{2}$As$_{2}$ below $\it T_{s,N}$ via measurements of detwinned 
single crystal, which allows us to measure single domain phonon structure in the AFM phase. 
The observed phonon anisotropy can be modeled by magnetic DFT calculations with a phenomenological reduction in 
force-constant anisotropy by roughly a factor of 3. In analogy to the small ordered moment in this materials, 
we suggest that the presence of magnetic fluctuations significantly reduces the phonon anisotropy that reflects
the coupling to the static magnetic order.

\begin{acknowledgments}
N.M acknowledges support from RIKEN Junior Research Associate Program. 
Work at SPring-8 was carried out under proposal numbers 2013A1467, 2013B1361, 2014A1207, 
2014B1760, 2015A1813. This work was partially supported by JST IRON-SEA project, Japan.
\end{acknowledgments}

\bibliography{bibliography}

\end{document}



\title{Supplemental Material for : Phonon anisotropy of detwinned {SrFe$_2$As$_2$} via inelastic x-ray scattering}


\author{N. Murai}
\affiliation{Materials Dynamics Laboratory, RIKEN SPring-8 Center}
\affiliation{Department of Physics, Osaka University}
\author{T. Fukuda}
\affiliation{Japan Atomic Energy Agency Quantum Beam Science Center (SPring-8/JAEA)}
\affiliation{Materials Dynamics Laboratory, RIKEN SPring-8 Center}
\author{T. Kobayashi}
\affiliation{Department of Physics, Osaka University}
\author{M. Nakajima}
\affiliation{Department of Physics, Osaka University}
\author{H. Uchiyama}
\affiliation{Japan Synchrotron Radiation Research Institute (SPring-8/JASRI)}
\affiliation{Materials Dynamics Laboratory, RIKEN SPring-8 Center}
\author{D. Ishikawa}
\affiliation{Japan Synchrotron Radiation Research Institute (SPring-8/JASRI)}
\affiliation{Materials Dynamics Laboratory, RIKEN SPring-8 Center}
\author{S. Tsutsui}
\affiliation{Japan Synchrotron Radiation Research Institute (SPring-8/JASRI)}
\author{H. Nakamura}
\affiliation{Center for Computational Science and e-Systems, Japan Atomic Energy Agency}
\author{M. Machida}
\affiliation{Center for Computational Science and e-Systems, Japan Atomic Energy Agency} 
\author{S. Miyasaka}
\affiliation{Department of Physics, Osaka University}
\author{S. Tajima}
\affiliation{Department of Physics, Osaka University}
\author{A. Q. R. Baron}
\affiliation{Materials Dynamics Laboratory, RIKEN SPring-8 Center}
\affiliation{Department of Physics, Osaka University}


\date{\today}


\maketitle

Here we consider a simple model meant to confirm expectations for the important time scales for a
vibrating system where there are fast fluctuations of the force constants.
We show that the relevant time scales are the mode splitting in the slow fluctuation limit,
which we call $\it s$, and the mean time between fluctuations, $\tau$.
The frequency broadening introduced by the fluctuations differs in two regimes determined by
\(\tau s << 1\) and \(\tau s >> 1\). The average mode frequency does not enter the problem. \\

We consider a mass, $\it m$, on a Hook's law spring \((F = ma = -kx)\)
where the force constant, $\it k$, switches between different values,
with the change assumed to be fast, and the dwell time for interval $\it i$ with constant
$\it k_{i}$, given by $\tau_{i}$ . Then on each time interval
\((t_{i}, t_{i+1}) = (t_{i}, t_{i} + \tau_{i}\)) one has a simple harmonic response
\begin{equation}
\label{eq:eq1}
x_{i}(t) = a_{i}\sin(\omega_{i}t + \phi_{i}) \hspace{3.8mm} (t_{i} < t < t_{i + 1}), 
\end{equation}
where \(\omega_{i} = \sqrt{k_{i}/m}\) and $\it a_{i}$, $\phi_{i}$ are given by 
boundary conditions.
We are interested in the power spectrum of the system, \(I(\omega) \equiv |\tilde{x}(\omega)|^{2}\),
where $\it \tilde{x}(\omega)$ is Fourier transform of the position given by 
\begin{widetext}
  \begin{eqnarray}
    \label{eq:eq2}
    \tilde{x}(\omega) & = & \int_{-\infty}^{+\infty}e^{-i\omega t}x(t)dt =
    \sum_{i}a_{i}\int_{t_{i}}^{t_{i+1}}e^{-i\omega t}\sin(\omega_{i}t+\phi_{i})dt \nonumber \\
    & = & \sum_{i}\frac{a_{i}}{\omega^2 - \omega_{i}^2}
    \Bigl[
      e^{-i\omega t_{i+1}}
      \Bigl(
      \omega_{i}\cos(\omega_{i}t_{i+1} + \phi_{i}) + i\omega\sin(\omega_{i}t_{i+1}+\phi_{i}) 
      \Bigr) \nonumber \\
      &&  \hspace{20mm}   - e^{-i\omega t_{i}}
      \Bigl(
      \omega_{i}\cos(\omega_{i}t_{i} + \phi_{i}) + i\omega\sin(\omega_{i}t_{i}+\phi_{i}) 
      \Bigr)
      \Bigr] \nonumber \\
      \end{eqnarray}
\end{widetext}

We limit ourselves to switching between two values, \(k_{2n+1} = k_{1}\) and \(k_{2n} = k_{2}\)
and assume \(k_{1} < k_{2}\), taking
\(\Omega = \omega_{2} - \omega_{1} = \sqrt{k_{2}/m} - \sqrt{k_{1}/m}\) and \(s = \Omega/2\pi\).
We assume the motion is continuous (there are no instantaneous translations of the mass)
and has fixed amplitude \(a_{i} = a \ \forall i\). The phases are then given by,
\(\phi_{i+1} = \phi_{i} + (\omega_{i} - \omega_{i+1})t_{i}\). 
We take the dwell times to be randomly distributed according to a negative exponential,
with mean dwell time, \(\tau\) ,
so that the probability of a particular dwell time is given by
\begin{equation}
\label{eq:eq3}
P(\tau_{i}) = \frac{1}{\tau}e^{\frac{-\tau_{i}}{\tau}} \\
\end{equation}

\begin{figure}[h]
\begin{center}
\includegraphics[width= 16.00cm]{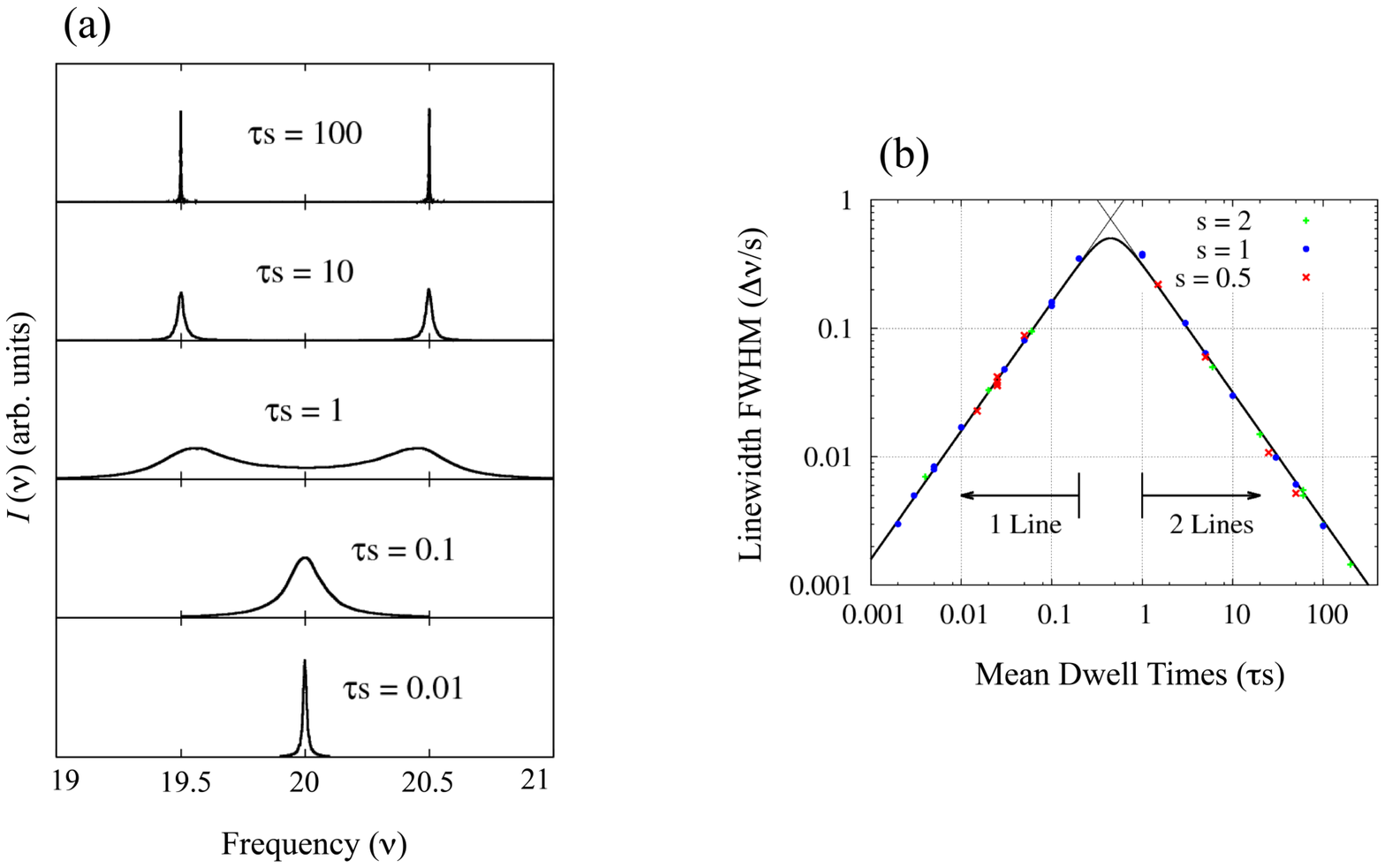}
\caption{(Color online)(a) Power spectra for different dwell times.
  The line splitting s was set constant at 1 and the frequency scale has arbitrarily been centered at 20.
(b) Normalized line width. The points are determined from the calculated power spectra
with splitting, s as indicated. The solid lines are the formulas from the text.}
\label{Fig1_s}
\end{center}
\end{figure}

Attempts to find a general closed form solution were not successful, but the problem can be
solved numerically by considering \(N\) intervals \((i=1...N)\), and letting \(N\) get large.
In fact one has to be slightly careful, as the Fourier transform, in the usual way,
will have fast oscillations (uninteresting noise) with a frequency scale given by
\(v_{osc} \sim (N\tau)^{-1}\) - one must insure that \(v_{osc}\) is small compared to
any line width in the problem. This forces \(N\) to be fairly large when $\tau$
becomes small. We also assume a finite frequency resolution, which also must be small
compared to any interesting line width.
Fig.\ref{Fig1_s} (a) shows the power spectrum \(I(\nu = \omega/2\pi\)) for several different values of the
mean dwell time. The progression from two well separated lines for \(\tau s >> 1\) to a
single collapsed line at the mean frequency for \(\tau s << 1\)
is clear, with a transition region with two broad overlapping peaks for \(0.2 < \tau s < 1\).
The line width, the full width at half maximum (FWHM), \(\Delta \nu\) behaves differently in each region.
Fitting the numerical results, for \(\tau s >> 1\)
one finds each peak has a width \(\Delta \nu \sim 1/(\pi \tau)\) while for
\(\tau s << 1\) the single peak is seen to have width approximately given by
\(\Delta \nu \sim \sqrt{8/\pi}\tau s^{2}\).
A formula that gives a reasonable fit (except near \(\tau s \sim 0.5\) where one
has two highly-overlapping lines) is
\begin{equation}
\label{eq:eq4}
\frac{\Delta \nu}{s} \sim \sqrt{\frac{8}{\pi}}\frac{\tau s}{\sqrt{1 + 8\pi \tau^{4} s^{4}}}
\end{equation}
As can be seen in Fig.\ref{Fig1_s} (b), this gives a reasonable approximation at least for
\(0.002 < \tau s < 200\), \(s = 0.5,1\), and \(2\).
We expect this to be valid over a larger range as well. 
